\documentclass[a4paper,11pt,pagebackref=true]{article}
\usepackage{amsmath,amsfonts}
\usepackage{algorithmic}
\usepackage{array}
\usepackage[utf8]{inputenc}
\usepackage[caption=false,font=normalsize,labelfont=sf,textfont=sf]{subfig}
\usepackage{textcomp}
\usepackage{stfloats}
\usepackage{url}
\usepackage{verbatim}
\usepackage{graphicx}
\usepackage{multirow}
\usepackage{hyperref}
\usepackage{caption} 
\usepackage{algorithmic}
\usepackage{bookmark} 

\hypersetup{   pdfborder={0 0 0},
              colorlinks = true,
              linkcolor=red,   
            linktoc=page,     
              filecolor=magenta,
              urlcolor=cyan,    
        citecolor=blue    
}

\usepackage{geometry}
\providecommand{\keywords}[1]
{
  \small	
  \textbf{\textit{Keywords---}} #1
}
\makeatletter
\usepackage{amssymb}
\usepackage{pifont}
\newcommand{\cmark}{\ding{51}}%
\newcommand{\xmark}{\ding{55}}%
\usepackage{fancyvrb} 
\newcommand{\ygg@basicalert}[2]{\fbox{\bfseries\sffamily\scriptsize#1}{
\sf\small$\blacktriangleright$\textit{#2}$\blacktriangleleft$}}
\newcommand{\annote}[2]{\ygg@basicalert{\textsc{#1}}{\textcolor{red}{#2}}}

\makeatother

\usepackage{balance}

\begin{document}
\title{Integrating Usage Control into Distributed Ledger Technology for Internet of Things Privacy\\
\large Preprint - To be published in IEEE Internet Of Things journal}
\author{Nathanaël Denis, Maryline Laurent, Sophie Chabridon}

\maketitle

\begin{abstract}
The Internet of Things brings new ways to collect privacy-sensitive data from billions of devices. Well-tailored distributed ledger technologies (DLTs) can provide high transaction processing capacities to IoT devices in a decentralized fashion. However, privacy aspects are often neglected or unsatisfying, with a focus mainly on performance and security.
In this paper, we introduce decentralized usage control mechanisms to empower IoT devices to control the data they generate. Usage control defines obligations, i.e., actions to be fulfilled to be granted access, and conditions on the system in addition to data dissemination control. The originality of this paper is to consider the usage control system as a component of distributed ledger networks, instead of an external tool. With this integration, both technologies work in synergy, benefiting their privacy, security and performance. We evaluated the performance improvements of integration using the IOTA technology, particularly suitable due to the participation of small devices in the consensus. The results of the tests on a private network show an approximate 90\% decrease of the time needed for the UCS to push a transaction and make its access decision in the integrated setting, regardless of the number of nodes in the network.
\end{abstract}


\section{Introduction}
\label{S_introduction}
The Internet of Things (IoT) is a unique paradigm, with an estimated 21 billion active devices -in 2021- connecting to and exchanging data through different kinds of communication networks~\cite{Jovanovic2021}. With a forecast number of active devices reaching 25.4 billion by 2030~\cite{Jovanovic2021}, the requirements regarding performance, security and privacy in the Internet of Things will be increasingly pressuring. The data generated by these numerous devices are often privacy-sensitive, due to their nature, e.g., health or financial data, or because they may infer and anticipate users' behavior. It is therefore crucial to protect these data with the proper tools.

According to the specific requirements of the IoT paradigm, distributed networks, in particular distributed ledger technologies (DLTs), have been actively studied as an appropriate solution. Decentralization is indeed a boon for the Internet of Things in many aspects. First, centralization limits scalability in terms of number of devices, as it increases deployment and maintenance costs compared to decentralized networks~\cite{Salimitari2020}. Centralization, often relying on cloud service providers, also affects security. Internal or external attacks as well as accidental disclosures can result in significant data leaks. Data availability is a major concern as well, put at risk by physical damages during natural disasters \cite{Lisheng2020} or more casually with denial-of-service attacks. Finally, cloud service providers often gather and analyze the users' data, which is an additional threat to privacy.

However, privacy is often not the priority for distributed ledgers and is not implemented by design~\cite{Wright2019}, even concerning the technologies dedicated to the IoT~\cite{Tennant2017}. Besides, the ledger is completely transparent in public blockchains, which creates a significant challenge in designing a satisfactory privacy solution. Some of the distributed ledger properties are legal conundrums, such as immutability which contravenes the 17th article of the European GDPR, the right to erasure~\cite{EUdataregulations2018}.

In that context, it is critical that users and devices can control the data they generate when relying on a distributed ledger. Usage control is an appropriate concept enabling fine-grain dynamic control over the data, with access based on \emph{authorizations}, \emph{obligations}, which have to be fulfilled to be granted access, and finally \emph{conditions} related to the system state. Modern usage control systems also monitor the data flow to prevent forbidden data dissemination. However, the usage control is not incorporated with the distributed ledger, adding potential bottlenecks due to interactions between the UCS and the distributed ledger nodes.
The purpose of this article is to answer the privacy and performance issues in large-scale IoT deployments, which can be summarized as the following requirements:

\begin{enumerate}
   \item \emph{Control over data}, which may translate into traceability as well as access and dissemination control;
   \item \emph{Scalability}, both in terms of participants and transaction processing capabilities for payments;
   \item \emph{No transaction fees}, large-scale IoT use cases will likely involve micro-payment scenarios.
\end{enumerate}

To address these requirements, this article proposes to integrate usage control with distributed ledgers based on directed acyclic graphs (DAG). The benefits of the integration are two-fold: 1) the components of the usage control system contribute to the security of the network, as a node; 2) the usage control system can process the transaction data without intermediaries, which is faster and more reliable.
We also provide an analysis of distributed ledgers based on their features, to determine which ledgers are relevant for the proposed integration. Finally, we present the integration of usage control into IOTA, a distributed ledger based on a directed acyclic graph. Performance tests are conducted on the implementation, showing approximately a 90\% decrease in the average needed time to push a transaction and take an access decision when the UCS is integrated.

This article is structured as follows. First, usage control and distributed ledgers are introduced in Section~\ref{S_background}, with a focus given on the use of distributed ledgers in the Internet of Things. Then, we discuss the related works in Section \ref{S_related_work}, and analyze why their limitations motivate this article. Section~\ref{S_integration}, after classifying the ledgers according to several parameters, details why an integration of usage control and distributed ledgers is beneficial and how to actually proceed to the integration. Appropriate use cases as well as an illustrative scenario are presented in the following Section~\ref{S_illustrative_scenario} to better clarify the expectations from integration. The performance evaluation of the integration is conducted in Section~\ref{S_performance_evaluation} using an IOTA-based implementation. Then, a privacy evaluation of usage control is provided in Section \ref{S_privacy_evaluation} before concluding in Section~\ref{S_conclusion}.

\section{Background}
\label{S_background}
In this section, we introduce the background about usage control, a privacy-enhancing technology to monitor the access and the dissemination of the data. Then, we introduce distributed ledgers, particularly blockchains and directed acyclic graphs, and how they benefit the Internet of Things.

\subsection{Usage Control}
\label{ss_usage_control}
Usage control can be considered as an extension of access control, with the purpose to continuously monitor data once access has been granted. It was first conceptualized by Park and Sandhu~\cite{Sandhu2003} as the UCON model. It introduces attribute mutability for continuous evaluation, as well as new decision factors: obligations and conditions. \emph{Obligations} are actions to be performed by the user to be granted access, either before or during access, respectively pre-obligation and ongoing-obligation. \emph{Conditions} are environmental and system-oriented decision factors that evaluate to true or false based on the system state. They can similarly be pre-conditions or ongoing-conditions. Conditions are not under direct control of individual subjects. An example of obligation would be to accept or reject cookies before accessing a website, while conditions may be the system load, the time of day or the security status of the system.
Another aspect of usage control is the ability to monitor the dissemination of information, known as \emph{information flow control} (IFC) or \emph{data flow control} (DFC)~\cite{Myers1997}. Modern usage control systems integrate flow control to avoid moving data outside the scope of the usage control system, or to transfer data to users who do not have the access rights~\cite{Fromm2020}. 

To monitor the access, the usage control system relies on policies to make its decisions. They are defined using a policy language such as XACML~\cite{Godik2003} or EPAL~\cite{Ashley2003}. Policies provide high-level rules and guidelines structuring who can access data and resources, and under which conditions.
 
\begin{figure}[t]
\centering
\includegraphics[width=0.55\textwidth]{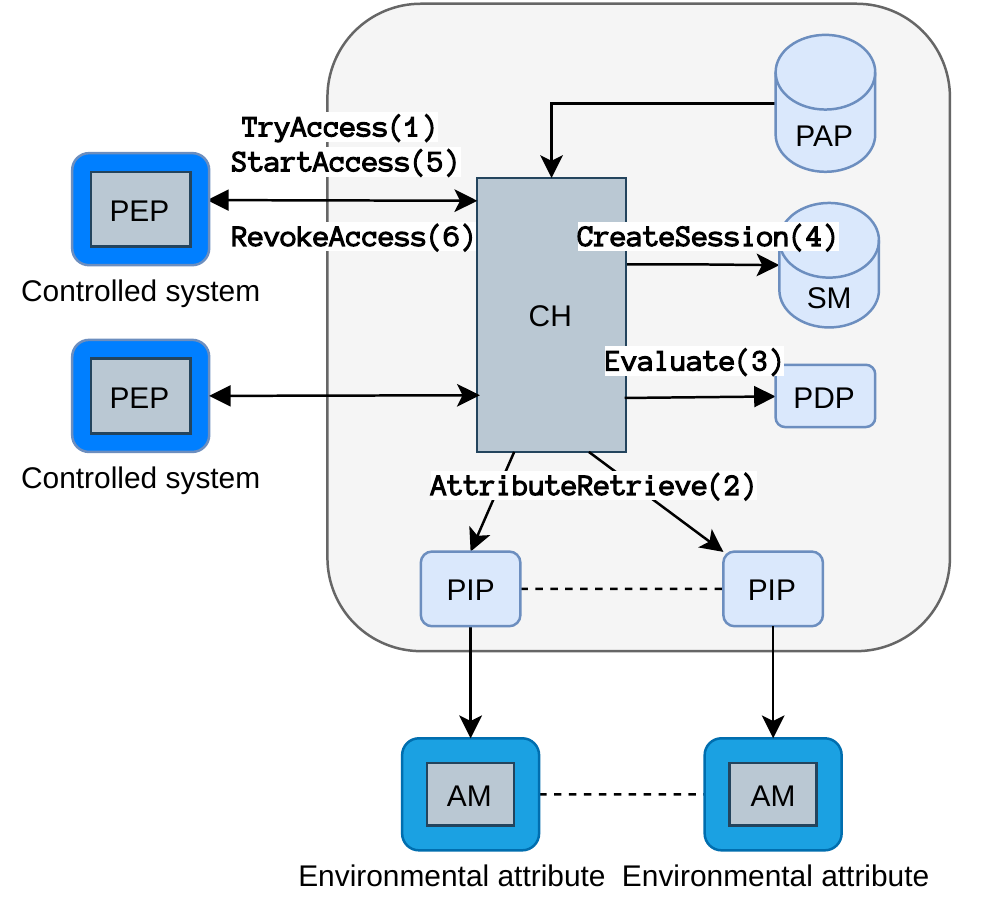}
\caption{Usage control framework, based on~\cite{Rizos2019}}
\label{F_ucon}
\end{figure}

The \emph{Usage Control System} (UCS) interacts with \emph{Environmental Attributes} through \emph{Attribute Managers} (AM) to recover the values of the attributes, and with the \emph{Controlled Systems}. It is composed of the following components, as shown in Figure~\ref{F_ucon}:
\begin{itemize}
   \item \emph{PDP}: the Policy Decision Point in charge of the policy evaluation. It takes as input an access request,
   the corresponding policy and the attributes of both users and context. Then it returns the result of the evaluation: \emph{Permit, Deny, Undetermined};
   \item \emph{PAP}: the Policy Administration Point, which stores the policies;
   \item \emph{PEP}: the Policy Enforcement Point enforces the policy evaluation result on a Controlled System;
   \item \emph{PIP}: the Policy Information Point, an interface so that the UCS can retrieve the values of the attributes from the system environment;
   \item \emph{CH}: the Context Handler, in charge of routing the different processes;
   \item \emph{SM}: the Session Manager stores all active sessions and the information needed for monitoring their status.
\end{itemize}

The workflow between the different components of the usage control system after an access is
requested is described in Figure \ref{F_ucon} and is composed of the following messages \cite{Rizos2019}. First, when a user requires access, the PEP sends a \texttt{TryAccess}(1) message to the CH. The CH requires the environmental attributes from the PIP (\texttt{AttributeRetrieve}(2)) and the policy from the PAP, and forwards them to the PDP for evaluation (\texttt{Evaluate}(3)). The PDP evaluates the access request to
\emph{PERMIT} or \emph{DENY}, and sends the decision to the CH which forwards it to the PEP. If the answer is \emph{PERMIT}, a unique SessionId is assigned to the access request (\texttt{createSession}(4)) and the Session Manager (SM) is updated. The PEP then sends a \texttt{startAccess}(5) message to the CH, to indicate the beginning of the access. If during the continuous evaluation, the UCS detects a policy violation, the CH informs both the PEP and the SM that the session is over by sending the \texttt{revokeAccess}(6) message. It is also possible that the user asks to end the session itself, by sending the \texttt{endAccess}(6') message. The CH tells the SM to delete session details, and the PIP to unsubscribe the attributes related to this session.

\subsection{Blockchains and Distributed Ledgers}
\label{ss_blockchains_distributed_ledgers}
A blockchain is a "distributed and immutable ledger made out of an unalterable sequence of blocks"~\cite{Salimitari2020}. There are two types of blockchains: public and private~\cite{Salimitari2020}. Public blockchains do not control access and are called permissionless, while private do have a control layer and are similarly called permissioned blockchains. Public blockchains are distributed and tamper-proof ledgers, not under the control of a single entity. Conversely, private blockchains require a third-party, usually a company, and their governance is centralized to some extent~\cite{Salimitari2020}.

To add a new block to a blockchain, participants of a blockchain network must agree on a \emph{consensus method}. The chosen method has a lot of consequences on the network features. There are several metrics to measure the performance of a blockchain network. \emph{Throughput}, generally expressed in a number of transactions per second (TPS), refers to how fast a blockchain can process transactions. \emph{Latency}, also known as block time, is the time between the addition of two blocks to the blockchain. Finally, \emph{scalability} is considered. It actually covers two different concepts~\cite{Steen2020}. First, scalability in terms of transaction processing capacity, corresponding to what the throughput measures. Then, scalability in terms of the number of users, which is positively impacted by the open membership, the lack of any centralized component and the absence of trust assumptions on any third-party~\cite{Steen2020}.

We now introduce the most used consensus methods and how they impact metrics:
The \emph{Proof of Work} (PoW) is a consensus method based on a computation race, where the first user solving the computation puzzle wins the right to add the next block to the ledger. The user appending the new block to the ledger is called the miner. The computation puzzle is most commonly based on a hash function. The users have to find a nonce solving the given problem, which requires a lot of computing power. Proof of work is well-known for its use by the Bitcoin cryptocurrency.
The \emph{Proof of Stake} (PoS) is the most used alternative to the proof of work for securing cryptocurrencies~\cite{Salimitari2020}. Contrary to the proof of work, the proof of stake is not based on an expensive computation race but the next miner is chosen at random based on its proportional stake in the network. The bigger the amount of cryptocurrency a user owns, the better the chances are to be chosen as the next miner. The proof of stake has a better latency and a slightly higher throughput compared to the proof of work \cite{Raghav2020};
The \emph{Proof of Authority} (PoA) is a variant of the proof of stake where the identities and the reputation of the nodes are at stake rather than a cryptocurrency asset. The time to reach the consensus and the latency are better compared to the proof of work, but not as good as the actual proof of stake~\cite{Raghav2020}.
In the \emph{Proof of Elapsed Time} (PoET), the miner is chosen at random based on a timer. The user whose timer expires first becomes the miner. This consensus method has several benefits, including a higher throughput and a low latency, i.e., the time between two blocks. However, its main drawback is its reliance on Intel's SGX, as the correctness of the timer execution must be verified within a \emph{trusted execution environment}, meaning that PoET's governance is centralized.
The \emph{Practical Byzantine Proof Tolerance} (PBFT) is a consensus method based on a vote. All the nodes are involved in the voting process and the consensus is reached when more than two thirds of the nodes agree upon the next block. As a consequence, the network can handle malicious behavior from at most a third of the nodes, which is low in comparison to the 51\% assumption in the proof of work networks~\cite{Aponte2021}. This renders the PBFT efficient in a private blockchain setting, but not for public blockchains which have a lower tolerance to malicious nodes~\cite{Salimitari2020}. The voting process does not scale well either, due to the network overhead it generates.

While blockchains are the most well-known instances of distributed ledgers for cryptocurrencies, the notion of \emph{Distributed Ledger Technology} (DLT) is wider and includes other technologies of interest. First, a distributed ledger can be completely disconnected from the notion of cryptocurrency, e.g., distributed databases. Besides, some cryptocurrencies do not build their transaction ledger using blockchains, but rather using different mathematical structures. The most used alternative to blockchain in cryptocurrencies is the \emph{directed acyclic graph} (DAG).

IOTA's Tangle~\cite{Popov2017} is built using a directed acyclic graph, as well as the Obyte ledger~\cite{Churyumov2017} and Nano~\cite{LeMahieu2017}. Though DAG-based technologies slightly differ, they share several interesting properties 
~\cite{Popov2017},~\cite{Churyumov2017}:
\begin{itemize}
   \item small transactions fees, or even no transaction fees at all;
   \item writing transaction is not energy-intensive;
   \item throughput is high after the bootstrapping stage and increases with the number of users;
   \item users add their transactions directly to the network without relying on miners or gateways.
\end{itemize}

\subsection{The Internet of Things and Distributed Ledgers}

Distributed ledgers, in particular blockchains, have not been originally thought for the Internet of Things. 
For instance, the Bitcoin network has a deliberately small throughput, i.e., number of transactions per second, to guarantee the security of the network and give the nodes enough time to verify the validity of the chain. A high throughput can be quantified as over 1000 transactions per second~\cite{Salimitari2020}, to be compared to Bitcoin's 6 transactions per second. Since the throughput required is much higher and the security requirements can be alleviated as transactions are not as substantial in value, the Internet of Things context is much different. Therefore, the research has been recently focusing on developing new kinds of distributed ledgers, often dedicated to the Internet of Things and taking variable forms.

Distributed ledger technologies provide several properties of interest for the Internet of Things. First, being at least partly decentralized, they alleviate deployment and maintenance cost compared to centralized networks~\cite{Salimitari2020}. Centralization also creates scalability and security issues. Second, they provide disintermediation and work even in the presence of a minority of malicious nodes, removing the necessity of having trust assumptions.

IOTA \cite{Popov2017} is a distributed ledger using a directed acyclic graph instead of a blockchain to represent the transactions, and it uses a light proof of work but to prevent spamming the network instead of securing the transactions. In this setting, the proof of work is no longer a computation race and is not energy
consuming. This methodology removes the bottleneck in blockchains, while matching the security needs of the IoT. Besides, when a user wishes to add a new transaction to the network, it has to validate two previous transactions beforehand. This validation method has by design the following outcome: as the number of users pushing new transactions increases, the pending transactions are validated faster. The network can therefore scale in terms of transactions.
The next version of IOTA, named IOTA 2.0, will switch the current consensus method to a leaderless voting-based consensus, removing the coordinator node, a centralized entity responsible for periodically validating groups of transactions. Besides, the proof of work is removed in IOTA 2.0, replaced by the \emph{rate control} mechanism. Rate control both prevents spams by replacing the PoW, and blocks the most heavily contributing nodes in overload scenarios, i.e., when transactions are too numerous for the network to handle them \cite{Popov2020}. Although the coordinator was critical for the network security during the bootstrapping phase, a benchmark from Wang \emph{et al.} \cite{Wang2022} indicates that milestones emitted by the coordinator do not significantly affect system robustness, while capping the transaction throughput. Other technologies based on directed acyclic graphs have been emerging, such as Obyte~\cite{Churyumov2017} which emphasizes the disintermediation and the very low transaction fees.

Private ledgers, such as the Hyperledger suite, can be helpful and bring several properties of interest: 1) as their name suggests, the data access is restricted to authorized users, improving privacy; 2) computation requirements are low and network response time is better. However, due to the centralization of governance, scalability is limited compared to public distributed ledgers, which in turn restricts the potential use cases \cite{Salimitari2020}.

\section{Related Work}
\label{S_related_work}

We will now introduce the related work about usage control and distributed ledgers, and how it addresses the requirements of large-scale Internet of Things deployments. The unaddressed requirements, scalability and free payments, are the motivation of our work.

\subsection{Usage Control and Distributed Ledgers}

Combining usage control with permissioned blockchains has been studied in the literature in particular for auditability, by writing the rules and the enforcement records on the ledger.

Khan \emph{et al.}~\cite{Khan2020} incorporated the components of the UCON framework into the Hyperledger Composer, to form the BlockU model. The authors use a private blockchain jointly with usage control to enforce usage control decisions automatically using smart contracts, and for the continuous monitoring provided by usage control, i.e., before, during and after the access. UCON is actually incorporated by using the Hyperledger Composer layer, a layer dedicated to modeling applications. Therefore, this incorporation can be seen as peripheral as the UCON components do not contribute to the blockchain network, and performance aspects are ignored. Besides, this work is not focused on IoT use cases.

Ma \emph{et al.}~\cite{Ma2020} proposed a decentralized usage control on both a public and a permissioned blockchain, respectively Ethereum and Hyperledger. A clear focus is given on privacy and decentralization to prevent accidental or intentional data leaks. All data operations are processed directly on the private blockchain with smart contracts, the degree of integration is consequently quite high. The public blockchain is used to introduce a reward mechanism, to encourage users to provide quality usage control data. However, the integration of usage control into the blockchain benefits usage control but not the blockchain network.

Shi \emph{et al.}~\cite{Shi2021} designed a Distributed Usage Control Enforcement (DUCE) solution for the Internet of Things, to tackle privacy issues related to the Cloud-Enabled Internet of Things (CEIoT). DUCE uses distributed PDP and PEP components and leverages private blockchains to store tamper-proof and traceable data on the ledger. The policies are written in the XACML language then converted by an algorithm into the Solidity language for smart contracts. DUCE relies on a \emph{trusted execution environment} (TEE) to enforce a trustworthy processing of the rules.

Zhang \emph{et al.} \cite{Zhang2022} devised an efficient data trading scheme with usage control for Industrial IoT. It is based on Hyperledger Fabric and relies on Intel SGX to define protected private regions of memory. Similarly to above-mentioned related works, the authors utilize smart contracts to automatically trade data. To tackle usage control overhead on the data trading system, the authors rely on \emph{Policy Monitors} (PM) on the user side, using the SGX as a TEE. This enables to offload usage policy enforcement on the user side, addressing scalability issues.

\subsection{Limitations}

As detailed in Section \ref{S_introduction}, the requirements regarding large-scale IoT deployments are 1) control over the data; 2) scalability; 3) transactions without fees.
All the related works, integrating usage control with a distributed ledger, solve the first requirement. However, the two remaining requirements are not addressed.

\textbf{Scalability.} The related works focus on private blockchains, which curbs scalability due to centralized governance. Besides, private blockchains relying on PBFT, notably used in both Ma \emph{et al.} and Khan \emph{et al.} articles with Hyperledger, can not be used for large-scale IoT networks due to network overheads~\cite{Salimitari2020}. Conversely, DUCE addresses scalability issues by distributing the Policy Decision Points and the Policy Enforcement Points. Performance tests show that their solution scales well in terms of \emph{usage control requests}, but does not consider blockchain transaction throughput.

\textbf{Transactions.} Hyperledger does not provide a cryptocurrency to make payments, which makes it unsuitable for IoT use cases involving transactions. Similarly, Shi \emph{et al.} only use distributed ledgers as it is decentralized, tamperproof, enables traceability and has mutual maintenance. It does not consider using the network for actual payments, but rather use the blockchain properties to preserve privacy. Zero-fee transactions are part of the requirements to enable micro-payments in the Internet of Things.

These two requirements must be addressed for large-scale IoT deployments. First, by identifying distributed ledgers with transaction capacities and scalable as regards throughput, and then by addressing scalability in terms of number of participants by improving usage control performance.

\section{Usage Control Integration}
\label{S_integration}
In this section, the principles of usage control integration with distributed ledger technologies are discussed. First, we identify several criteria to determine if a technology is suitable for integration in Section~\ref{ss_integrability_criteria}. Then Section~\ref{ss_integration_benefits} details the expected benefits of integration.
Finally, Section~\ref{ss_integration_methodology} describes the methods to integrate the UCS into distributed ledgers.

\subsection{Integrability Criteria}
\label{ss_integrability_criteria}

The distributed ledgers are very diverse in nature, due to their consensus methods, incentives, access control methods or even transaction ledgers. In order to design a relevant integration of a usage control system, it is first necessary to identify the goals of this integration and to find out the suitable distributed ledger technologies. We therefore propose a classification (cf. Table~\ref{tab:classification}) relying on three categories of features: the consensus method, the incentive to contribute to the network, and the transaction ledger.

\begin{figure}[t]
\centering
\includegraphics[width=0.65\textwidth]{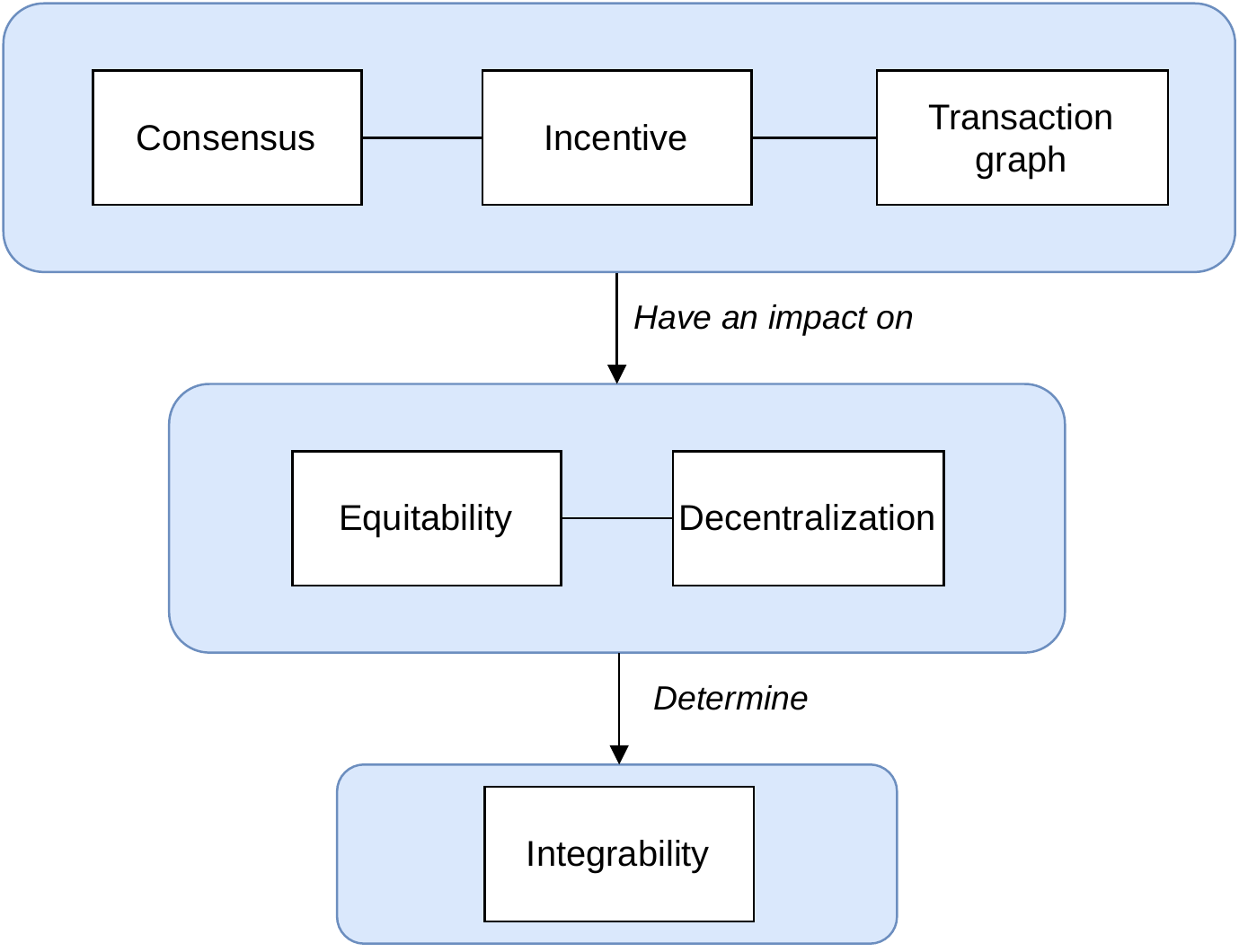}
\caption{Criteria used for integration suitability - schematized}
\label{F_analysis}
\end{figure}

\textbf{Method of analysis.}
To determine whether a technology can integrate properly with the usage control system, we consider two properties of distributed ledger features: decentralization and equitability. These properties are directly influenced by other ledger features and are the basis of our analysis: 1) the consensus method; 2) the incentive to contribute to the network; 3) the transaction ledger structure. The method of analysis is schematized on Figure \ref{F_analysis}.

\emph{Equitability} describes the possibility for every user to have a fair part in the decision-making and consensus processes. Particularly, it includes devices with poor computation or storage capacities. This property ensures small devices in the network have an actual impact, and that a small fraction of users do not monopolize the power in the network.
\emph{Decentralization}, besides providing performance and security benefits, is an interesting property to assess the integration suitability. Indeed, decentralization partially reflects the above-mentioned equitability. A fully decentralized network is likely to rely on local consensus to make decisions, which gives a bigger impact to participating devices~\cite{Steen2020, Popov2020}. 
Decentralization can take several forms, all expressing power asymmetries~\cite{Bodo2021}:  
1) decentralization in the governance; 2) decentralization in terms of power, where nodes, or more likely node aggregates, have a disproportionate power over the rest of the network. For instance, the proof of work consensus method gives power to the most powerful mining pools, while the proof of stake creates another form of centralized capital power. Actually, a highly centralized governance is often a prerequisite for a distributed network to work, which means a network can be highly centralized regarding only one centralization aspect, i.e., governance or power, but not the other.
The integration is more relevant as fairness and decentralization increase.
The distributed ledger features have an impact on these two properties of equitability and decentralization. To determine the characteristics that favor or hinder integration, three categories of features are next analyzed.

\textbf{Consensus method.}
Just like distributed ledgers, consensus methods are very diverse but will be summarized as:

\begin{itemize}
   \item \emph{Proof of Work}: due to computation concerns, a challenging computation race will exclude low-power devices from participating in the consensus. Even though the UCS could be used to make a profit while fulfilling its assignments, it is not an interesting application for our integration project;
   \item  \emph{Proof of Stake}: PoS better fits the IoT computational requirements, but has several drawbacks for integration. First, you can not contribute to the consensus unless you have a stake. Many devices, particularly sensors, do not have any reason to hold a cryptocurrency asset, either for the consensus or to make transactions with it~\cite{Raghav2020}. The proof of stake also tends to centralize the network in the hands of a few users, either the delegates (DPoS) or the richest users. Equitability is therefore not achieved with this consensus method. Integrating the UCS in a distributed ledger based on PoS is therefore not relevant, as they can not contribute to the network;
   \item \emph{Proof of Authority}: PoA, contrary to the proof of stake, does not require holding a cryptocurrency asset. Nevertheless, PoA, by delegating the power to a few nodes considered trustworthy, limits the possibility to integrate the UCS components;
   \item \emph{Proof of Elapsed Time}: in PoET, miners are chosen at random using timers. This consensus method therefore guarantees equitability. In this setting, the components of the usage control system could contribute to the consensus, and integration is relevant. However, PoET is limited to private ledgers, as users must first join the network and gain a membership certificate to be allowed to start the timer;
    \item \emph{Practical Byzantine Fault Tolerance}: all nodes take part in the voting process with an equal power, providing the equitability property. Besides, this consensus method is suitable for the IoT but only on private ledgers due to scalability issues in terms of the number of users, as it causes high network overhead~\cite{Salimitari2020}.
\end{itemize}

Relying only on consensus methods, it appears that only the practical byzantine fault tolerance (PBFT) and the proof of elapsed time (PoET) are suitable for integration, but are used only in private settings. This explains why the related works (cf. Section \ref{S_related_work}) mostly focus on private ledgers, when the requirements for the IoT use cases are not considered.

\textbf{Incentive.}
Most consensus methods rely on a financial incentive to encourage users to contribute to the network. The proof of work rewards the miner when it adds a block to the ledger, while the proof of stake rewards the users when they stake their cryptocurrencies for network security. However, rewards have negative side effects. When using a proof of work, it not only encourages miners to contribute to the network, but also to group together in mining pools. This phenomenon is the cause of the centralization of the network and users can no longer act independently~\cite{Ketsdever2019}. In proof of stake networks, the staking reward incentives create the \emph{Nothing at Stake} issue. When a fork occurs, i.e., when two versions of the ledger are competing, the validators have an interest in maintaining both versions to avoid taking the risk of maintaining the wrong one, and earn worthless rewards~\cite{Ketsdever2019}.

In the Internet of Things context, where small devices with energy constraints are involved, the contribution to the network may not be conditioned by financial incentives. IoT-oriented projects would rather focus on operational and energy savings, in particular for battery-powered devices. This is the case for IOTA, which does not introduce any kind of financial rewards for operating a node or validating transactions. The usage control system, as a security device, does not contribute to the network for financial rewards, but seeks to deliver fast access decisions at the minimum cost.

\textbf{Ledger type.} Considering consensus methods, e.g., PoET and PBFT, it appears that private blockchains are more suitable for integration than public blockchain. However, directed acyclic graphs have several properties of interest, among which some significantly ease the integration process. The removal of gateways in directed acyclic graphs enables the users to push their transactions directly to the network. This is also true for the usage control system, which may actively contribute to the network. The overall metrics of directed acyclic graphs enable more users to contribute to the network. The usage control system can make its decisions faster by processing the transactions locally.

\textbf{Selection of the suitable distributed ledgers.}
To sum up, the possibility to integrate usage control with distributed ledgers is mostly determined by the equitability of the protocol and its decentralization. Three main criteria have a direct impact on equitability and decentralization: the consensus method, the incentive and the transaction ledger.

The consensus method has a direct impact on the possibility for small devices to contribute to the network. Incentive is paramount as well because usual financial incentives tend to centralize the network. Finally, the ledger type, either a blockchain or a directed acyclic graph, has a deep impact on the network features. Since DAGs are meant to allow users to push and check transactions themselves, they enable the contribution of the usage control system to the network.

 According to this classification, summarized in Table~\ref{tab:classification}, it is possible to depict two categories of fitting technologies, both without a financial incentive: directed acyclic graphs and private blockchains. However, only directed acyclic graphs consider the large-scale IoT requirements, while private blockchains are not scalable and do not have cryptocurrencies. Consequently, \emph{we will consider directed acyclic graphs} for the integration of usage control in the following.
\begin{table*}
\begin{center}
\begin{tiny}
\setlength{\extrarowheight}{5pt}
\begin{tabular}{ |c|c|c|c|c|c|c| }
\hline
Parameter & Instance & IoT Suitable & Equitability & Decentralized* & Integration & Notes\\
\hline
\multirow{7}{4em}{Consensus} & PoW & \xmark & \xmark  & \xmark (.pow) & \xmark & Compute-intensive\\

& PoS & \xmark & \xmark & \xmark (.pow) & \xmark & Stake needed\\

& PoA & \xmark & \xmark & \xmark & \xmark & Similar to PoS \\

& PBFT & \cmark & \cmark & \cmark & \cmark & Private blockchains only \\

& PoET & \cmark & \cmark & \xmark (.gov) & \cmark & Private blockchains, relies on Intel SGX\\
\hline
\multirow{3}{4em}{Incentive} & Financial & \xmark & \xmark & \xmark & \xmark & Deterrence for small devices \\
& Savings & \cmark & \cmark & ? & \cmark & Incentive for small devices\\
\hline
\multirow{3}{4em}{Ledger} & Private blockchain & \cmark & \cmark & \xmark (.gov) & \cmark & Promising but scalability, currency issues\\
& Public blockchain & ? & ? & ? & ? &  Very diverse, not determining \\
& Directed acyclic & \cmark & \cmark & \cmark & \cmark & Significantly favors integration\\
\hline
\end{tabular}
\captionof{table}{DLT parameters and their impact on integration. Question marks (?) mean the parameter is not determining. \\ *Decentralized can take several forms, governance (.gov) or power asymmetries (.pow)}
\label{tab:classification}
\end{tiny}
\end{center}
\end{table*}

\subsection{Integration Benefits}
\label{ss_integration_benefits}

A logical way to integrate the usage control system with a distributed ledger based on a DAG is to run a node. Nodes are indeed critical components of distributed ledgers. They differ depending on the technology, but their purpose is at least to check the validity of transactions. The usage control system could run a node with the following expected benefits for itself:

\begin{itemize}
   \item \emph{disintermediation}: running a node avoids relying on a third-party. The bandwidth is secured for its own transactions without delay, making the transaction-based usage control more efficient;
   \item \emph{storage control}: it is possible to keep all the transaction records needed to enforce the policies. Nodes may indeed rely on local snapshots to reduce the size of the local ledger;
   \item \emph{node configuration}: it is possible to configure the node and adjust both security and performance parameters to the UCS needs;
   \item \emph{network security}: the UCS will contribute to the ledger verification, reducing the probability of failure resulting from a low number of nodes \cite{Khacef2021}. Besides, a higher number of nodes makes some attacks harder, such as the 34\% attack, which is the DAG equivalent of the blockchain 51\% attack \cite{Ullah2021,Aponte2021};
   \item \emph{throughput increased}: the UCS will increase the throughput as it will push transactions on the network, due to DAG properties;
\end{itemize}


To fulfill its mission, the usage control system has to monitor system and network calls, which is an intrusive process. \emph{Transparency} and \emph{auditability} are paramount in this context.
Transparency in usage control can be considered as the fact that the usage control operations, e.g., allowing access or preventing dissemination, are communicated to the others, while auditability anticipates the storage of these usage control data for a future audit.
Transparency and auditability can be achieved by writing usage control data on the ledger such as the operations performed by the UCS on the users. Auditability reinforces the trust in the usage control system: 1) for data owners, as it is possible to check that the usage control of their data is indeed enforced, and by which means; 2) for monitored users requiring data access, as any intrusive misbehavior of the usage control system will be publicly reported.

\subsection{Integration Methodology}
\label{ss_integration_methodology}

The integration methodology concerns two aspects: the global system architecture, considering how the usage control system can integrate the distributed network, and data protection, as data written on the ledger becomes visible and exposed to privacy risks.

\begin{figure*}[t]
\centering
\includegraphics[width=\textwidth]{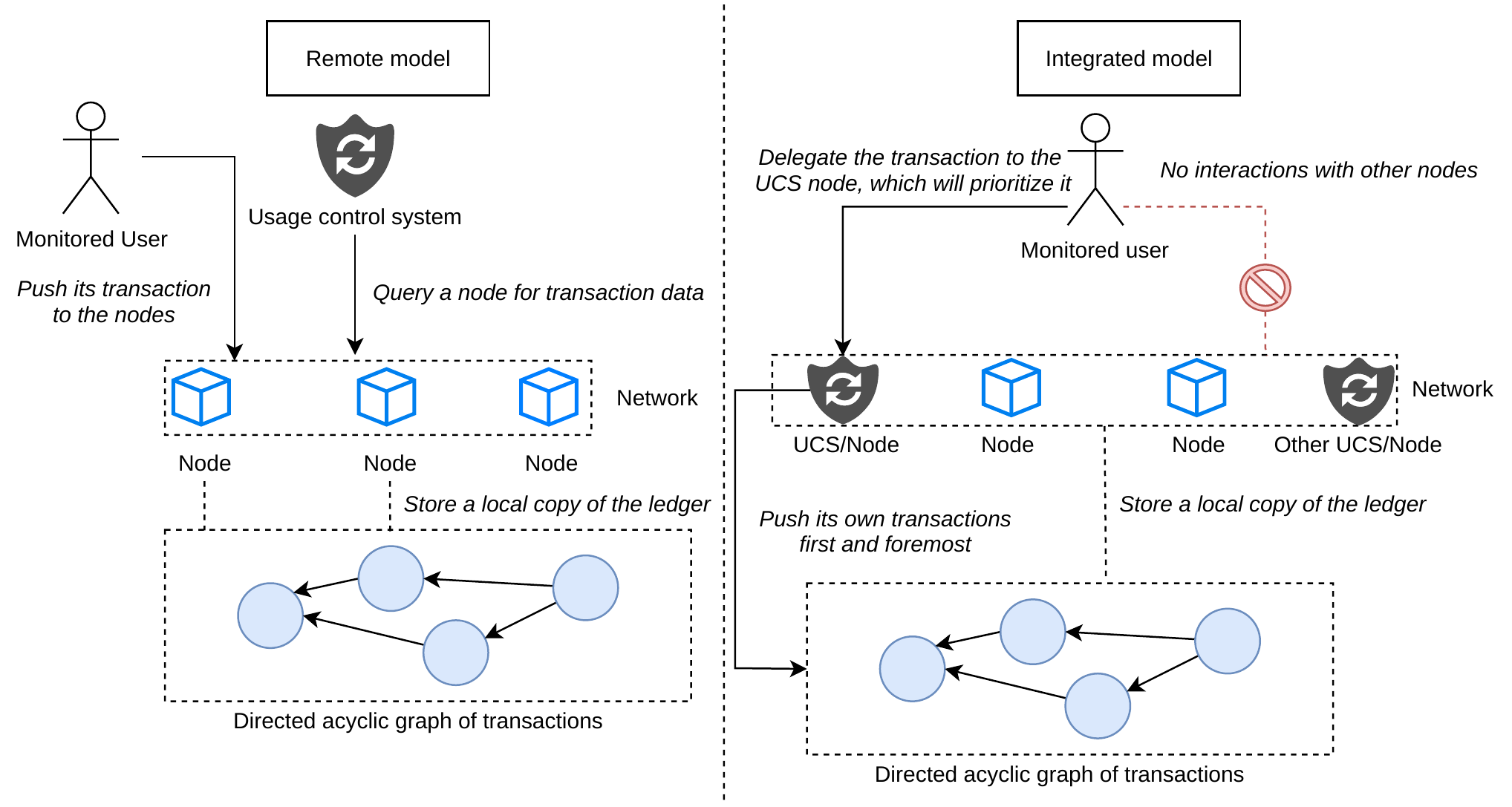}
\caption{Differences between the remote model and the integrated model when using directed acyclic graphs.}
\label{F_peripheral_vs_integrated}
\end{figure*}

\textbf{Integration as a Node.}
Directed acyclic graphs, by design, alleviate the requirements on nodes in order to maximize the number of devices contributing to the network. Therefore, the usage control system itself can deploy a node and push transactions itself. By deploying a node, the usage control system will 1) push transactions including some related to usage control use cases, accelerating the access decision process; 2) prioritize its own transactions on the network, avoiding a potential queue; 3) store a local copy of the ledger, to process the ledger faster when necessary for its access decision. The integration model is represented in Figure~\ref{F_peripheral_vs_integrated}, showing the links between the UCS, the nodes, and the monitored users. Data can be partially written on the public ledger as long as they do not create inference risks, as described in the following.

\textbf{Data management.}
For public ledgers, one key motivation for integration is the capacity to write immutable data on the ledger for transparency purposes (cf. Section~\ref{ss_integration_benefits}). Data must be classified to determine whether they can be displayed publicly or not. Data can be of four types, summarized in Table~\ref{tab:storage}:

\begin{itemize}
   \item \emph{protected data}: data whose access is monitored by the UCS;
   \item \emph{usage control data}: data concerning the usage control, including the processes performed by the UCS as well as the results of the evaluation;
   \item \emph{users' data}: data needed by the UCS about the users, such as their attributes to make access decisions;
   \item \emph{metadata}: data about the other data, e.g., data that states the kind of processed attributes, but not the content of the attributes.
\end{itemize}

\emph{Protected data} must not be stored in clear text on the ledger, and can either be encrypted on the public distributed ledger or stored in a distributed database. However, encryption produces computational overhead and the management of encryption keys is a real issue in large scale IoT contexts; that is why we resort to using a distributed database and exclude encryption. The purpose of the database is to store the protected data by restricting its access, and to interact with the usage control system, which decides if a user can be granted access (cf. Figure \ref{F_system_model}). \emph{Usage control data} describe the operations performed by the usage control system on the users. They are basically composed of a data identifier, the pseudonyms of the users and the action performed. They are published on the public ledger for transparency and auditability purposes. \emph{Data about users}, such as their attributes, are needed only by the usage control system and are stored in the database. Finally, \emph{metadata} is published on the ledger, such as timestamps and data identifiers to keep track of the data. Metadata can pose a \emph{detectability threat} when revealing its existence can lead to privacy issues, even without providing the actual content of the data, e.g., knowing the existence of a police record associated with an identity, without the content inside the record, is a sensitive information. Metadata must therefore be processed carefully, using a privacy threat analysis framework such as LINDDUN~\cite{Deng2011}. Such a privacy threat analysis is conducted in Section \ref{S_privacy_evaluation} with an illustrative scenario.

\begin{table}[htbp]
\centering
\begin{tabular}{ |c|c| }
\hline
Data type & Data storage \\
\hline
Protected & Database \\
Users' data & Database \\
Usage control & Directed acyclic graph \\
Metadata & Directed acyclic graph*\\
\hline
\end{tabular}
\vspace*{1mm} 
\caption{Data types and their respective storage area \\ * If detectability is not an issue}
\label{tab:storage}
\end{table}

\section{Illustrative Scenario}
\label{S_illustrative_scenario}
To emphasize the usefulness of integration, this section first gives the general use cases where integration is expected to be beneficial, before giving a more precise illustrative scenario about electricity consumption prediction. The system model according to this scenario is presented next.

\subsection{Use Cases}

The benefits of integration, as described in Section~\ref{ss_integration_benefits}, are related to performance gains when the transactions are involved in the usage control. Therefore, the use cases which will benefit the most from integration are those involving buying or selling data. Advertising and marketing in general are use cases where actors want to sell data to others.
Besides, use cases related to the IoT are even more relevant as devices generate valuable data in a substantial quantity. Owners of devices have a strong incentive to sell the collected data, but in a privacy-preserving way as the data they generate are privacy-sensitive, e.g., location data create risks of inference. Usage control, including dissemination, is paramount for these use cases.

As an illustrative scenario, we consider an electric utility company that wants to predict the electricity demand peaks due to the use of air-conditioners. For that, the company relies on both temperature and humidity levels in several areas, and predicts the use of air-conditioners according to these parameters. However, the electric utility company does not want to deploy sensors itself as it is much more expensive than buying the data from data providers. The devices are numerous and can dynamically join or quit the network.

As the location of the data is given along the temperature and humidity levels, there is a re-identification risk for the data provider as location is very privacy-sensitive. Therefore, the data provider wants to control the data usage of the company, to prevent both dissemination and inference analysis. Besides, the data provider wants to give the access only after the payment has been made. For these reasons, it is necessary to implement usage control. In order to make the data available for sale, the data provider could either:

\begin{itemize}
   \item rely on smart contracts on a private blockchain, to grant data access once all requirements are met;
   \item rely on a directed acyclic graph and check as a node that the transactions are completed, before granting access.
\end{itemize}

However, in this large-scale setting, the private blockchain will not scale well. The data must be bought from a third party, requiring the use of a cryptocurrency as well.
The electric utility company will therefore opt for the directed acyclic graph to benefit from decentralization, scalability and the ability to make free payments.

\subsection{System Model}

\begin{figure*}[t]
\centering
\includegraphics[width=\textwidth]{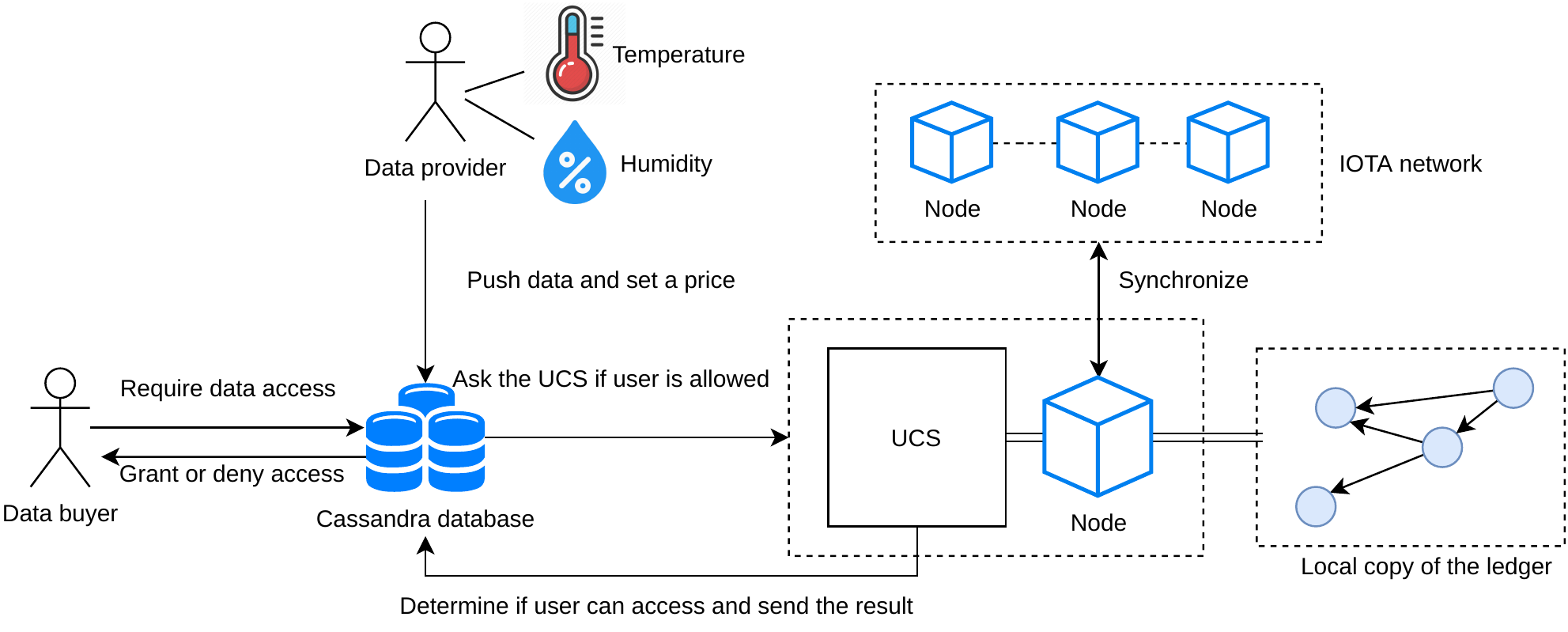}
\caption{Interactions within the system model during the pre-access - the electricity consumption prediction scenario}
\label{F_system_model}
\end{figure*}

Based on the electricity consumption prediction scenario,
the agents of the system can be summarized as follows. First, the \emph{data providers} that sell temperature and humidity data collected from their devices. Then the electricity utility companies, referred to as the \emph{data buyers}, buy the temperature and humidity data from data providers to predict the electricity demand. The \emph{usage control system} (UCS) is responsible for monitoring the access rights of the companies and for preventing dissemination, based on their attributes and their actions. In particular, it processes the transactions on the \emph{distributed ledger} to grant access if the payments are performed. A \emph{distributed database}, shared between the data providers, stores the protected data. \emph{Network nodes} validate the data buyers' transactions and propagate them to the other network nodes. The system model is depicted in Figure~\ref{F_system_model}.

\subsection{Interactions}

\begin{figure*}[pt]
\centering
 \includegraphics[width=0.9\textwidth]{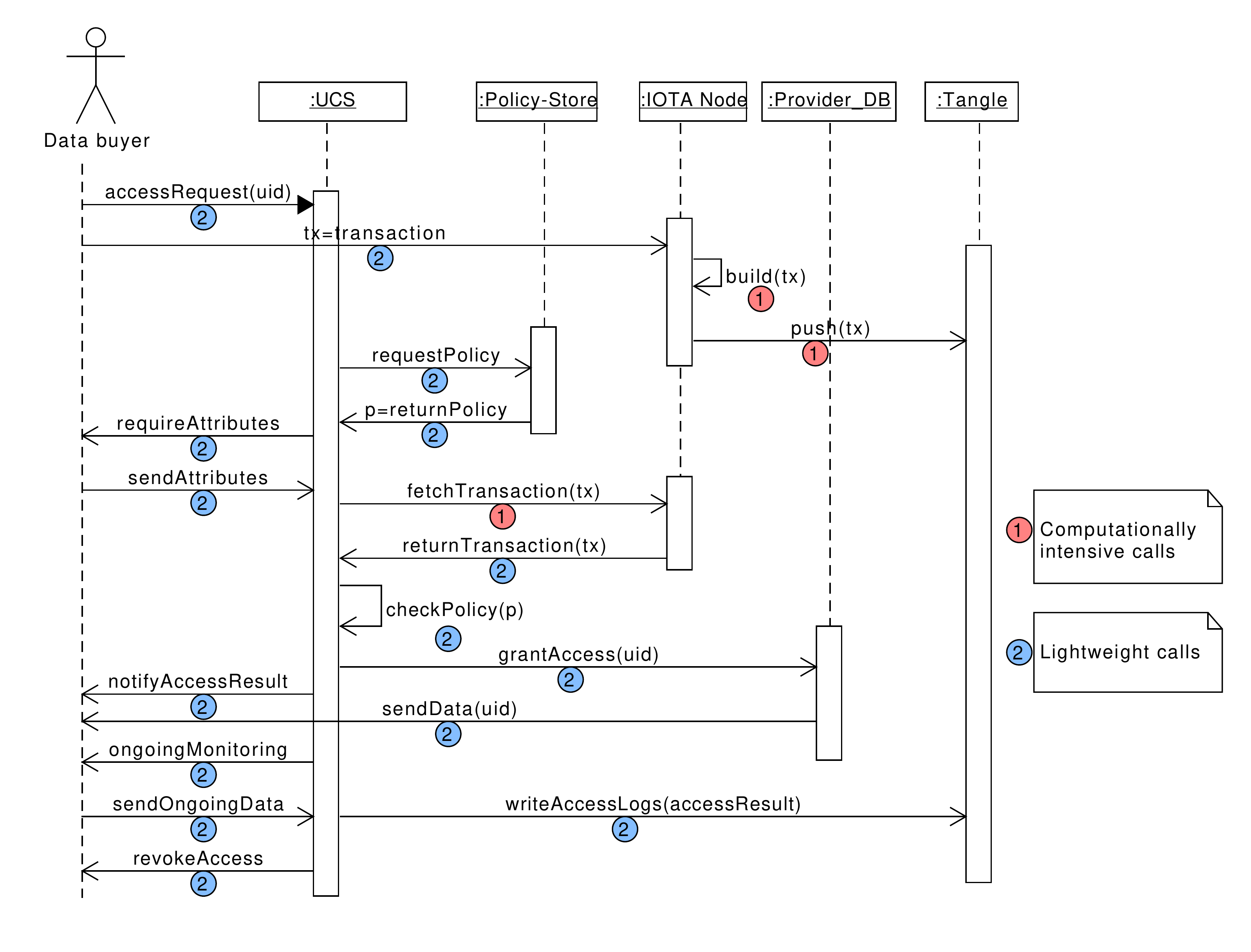}
\caption{Sequence diagram for pre-access and ongoing access - UCS not integrated with IOTA}
\label{F_sequence}
\end{figure*}

In this section, the system and network calls granting or denying an access are detailed and classified into two categories according to the intensity of computations required, as depicted in the sequence diagram of Figure \ref{F_sequence}.
First, the data buyer initiates the access by sending an \texttt{accessRequest} with its user identifier. Then it sends a \texttt{transaction} to perform the payment, which is processed by an IOTA node. This node first checks if the transaction is legitimate by checking it against the local state of the ledger. If the transaction is indeed legitimate, the node computes a light proof of work only to prevent spam, and not as part of a consensus method. The node then propagates the transaction to the rest of the network through its peers using the \texttt{push} call. The UCS can start monitoring the access and therefore sends a \texttt{requestPolicy} to its policy store, to fetch the XACML policy specified by the data provider beforehand. The UCS then asks attribute values from the data buyer (\texttt{requireAttributes}) to check the policy compliance. The attribute values will be used for data access control, but also include values resulting from network calls analysis for information flow control. Once the UCS receives the attributes (\texttt{sendAttributes}), it fetches the transaction status on the network by calling \texttt{fetchTransaction}, then determines the policy compliance ( \texttt{checkPolicy}) to make an access decision. If the user is authorized to access the data, i.e., UCON authorizations, obligations and conditions are fulfilled, the UCS sends a \texttt{ grantAccess} call to the data provider and notifies the data buyer with a \texttt{notifyAccessResult}. In response to the \texttt{grantAccess} call, the data provider sends the purchased data to the data buyer with the \texttt{sendData} call.
The data buyer is monitored by the UCS for the ongoing obligations and ongoing conditions (cf. Section \ref{ss_usage_control}) with usually multiple \texttt{ongoingMonitoring} calls, requiring environmental attributes. The monitored data buyer must reply to \texttt{ongoingMonitoring} with \texttt{sendOngoingData}, so that the UCS can make its access decisions and react if the data is disseminated. The UCS writes usage control logs and metadata for transparency while considering privacy threats (cf. Section \ref{ss_integration_methodology} and Table \ref{tab:storage}) by calling \texttt{writeAccessLogs}. The access is eventually revoked by the UCS with \texttt{revokeAccess}, should the data user end it willfully or
by contravening the data provider's policy.

\section{Performance Evaluation}
\label{S_performance_evaluation}

This section details an instance of a distributed ledger based on a directed acyclic graph, which is integrated with a usage control system for privacy purposes. Following the illustrative energy consumption prediction scenario, the purpose of the evaluation is to show that the integration is efficient and to assess its estimated outcomes in terms of performance. To this end, we will measure the time needed for the UCS to make an access decision in the integrated setting compared to the remote node setting.

\subsection{Testbed}
\label{ss_testbed}
\begin{figure}[p]
\centering
\includegraphics[width=0.7\textwidth]{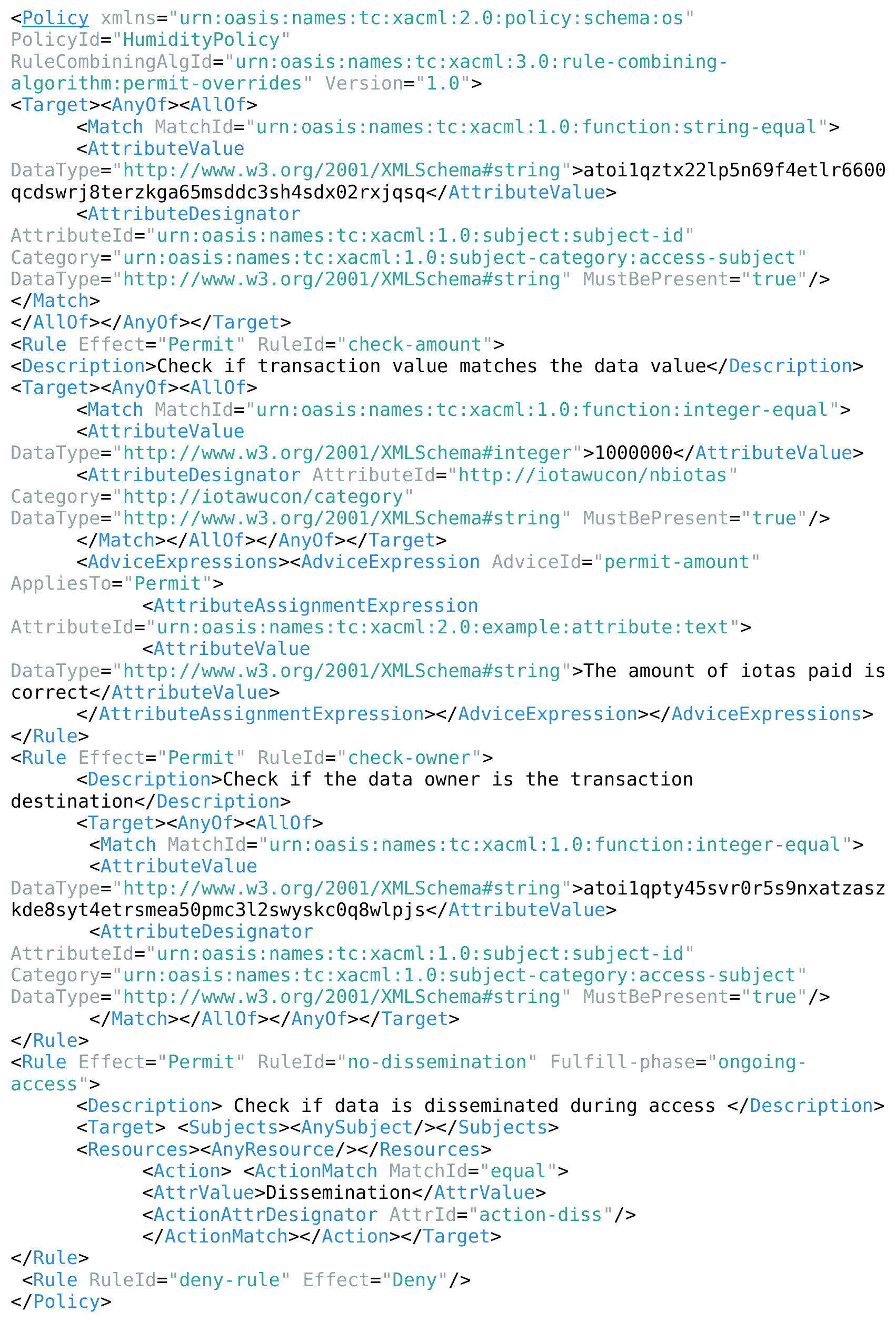}
\caption{Example of XACML policy at the UCS. Through the rules defined by the user, the data buyer is granted access to humidity data after payment is made (UCON pre-condition) and the dissemination is monitored during the access (ongoing obligation).}
\label{F_xacml_policy}
\end{figure}

 \textbf{Testing environment.} To assess our contribution, we test our integration using the IOTA technology, for two reasons: 1) it is the most studied DAG-based technology in the literature; 2) it focuses on IoT use cases. Free transactions are also a plus; the small transaction fees on other DAG-based technology have an impact when processing a lot of transactions simultaneously, which is expected regarding the use case. Moreover, since data is partially stored off-chain (cf. Section~\ref{ss_integration_methodology}), we relied on a \emph{Cassandra} distributed NoSQL database as a decentralized storage. On top of the distributed aspects, Cassandra is horizontally scalable which means it can easily handle increasing traffic demands by adding more machines~\cite{Silva2021}. Cassandra can also work on a low-power cluster making it particularly suitable for the Internet of Things~\cite{Silva2021}.
 The node is powered by \emph{Hornet}, a community-driven IOTA node software written in the language Go. The usage control system is written in Java, and the interaction with the Tangle is managed using the Rust library bindings for Java. The usage control policies are defined by the users and written using the XACML language~\cite{Godik2003} as in the example given in Figure \ref{F_xacml_policy}. During the tests, policies are not specified by users, but automatically derived for convenience.

\textbf{Network selection.}
IOTA has a public development network called \emph{devnet} for testing. Contrary to the mainnet, the devnet has free tokens and is meant for testing. The devnet could be convenient because its network is already deployed and easy to use. However, it has major drawbacks for testing. First, the network is subject to complete overhauls with no backward compatibility, preventing reproduction of the tests conducted in our experiments on the same network. In particular, the introduction of IOTA 2.0 will likely lead to the removal of the current testing network, designed for the first version of IOTA. Second, as the network is public, it is not possible to control the network topology, e.g., the number of nodes.

\begin{figure}[tp]
      \centering
      \includegraphics[width=0.7\textwidth]{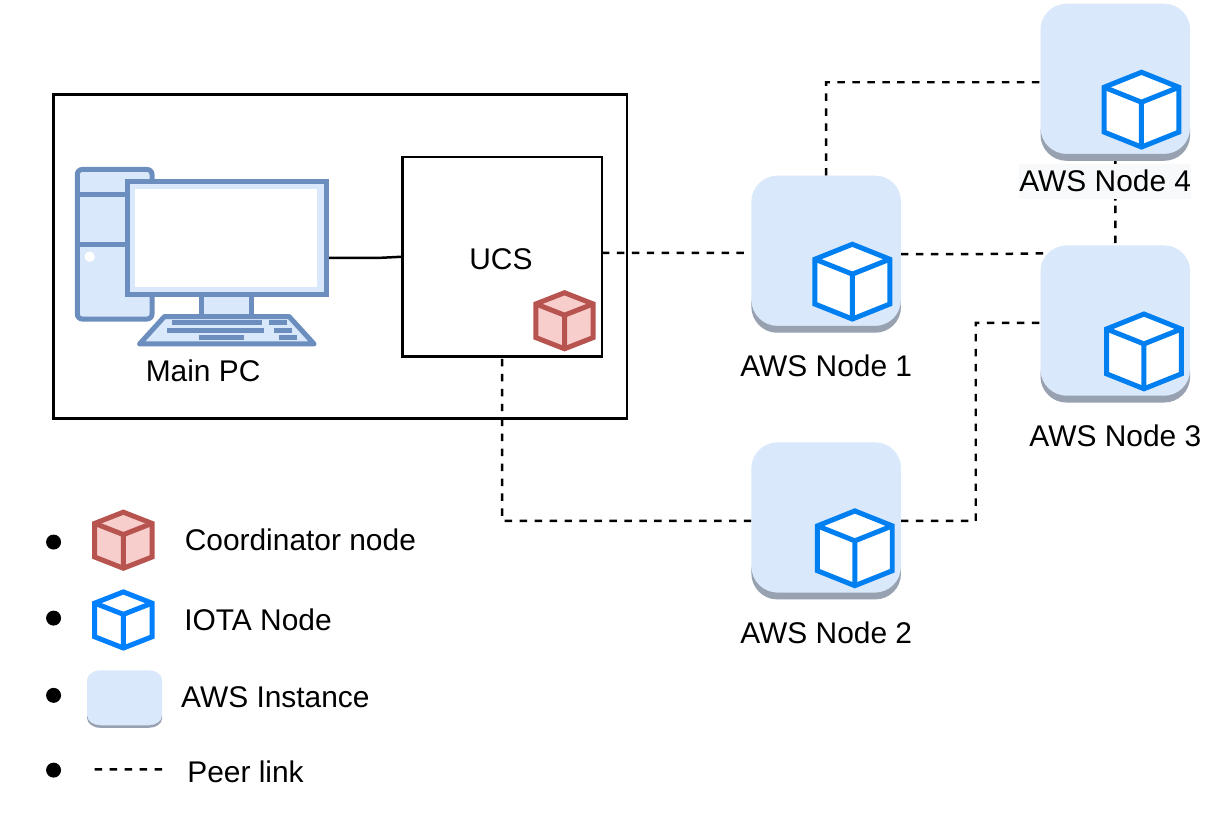}
      \caption{Private Tangle architecture with AWS instances - 5 nodes. Each instance runs an IOTA node, and a computer runs the Coordinator node and orchestrates the network.}
      \label{F_network_architecture}
      \end{figure}

The tests on the public network alone are not sufficient to demonstrate the efficiency of integration.
To ensure our tests can be reproduced, we instead deploy a \emph{private Tangle}. This methodology has been used by Dong \emph{et al.} \cite{Dong2019} to benchmark different DAGs. Several nodes deployed on the AWS instances will constitute a private IOTA network. The network architecture of the private Tangle with AWS instances is given in Figure \ref{F_network_architecture}, for 5 nodes.

\textbf{Nodes settings.} The tests are conducted using AWS t2.micro instances with 8 Gio storage capacity, 1 vCPU and 1 Gio RAM, in accordance with the Internet of Things constraints. An additional device is used to run the usage control system, which has more storage and computational power: 8Gio of RAM, 32Gio of storage and 4 CPUs. The nodes are located in Amazon's default US East (North Virginia) zone. Each Hornet node on the private Tangle uses the default spammer, spamming each 5 messages by second. Spamming is a desired behavior, as it theoretically speeds up the transaction validation in IOTA.

\subsection{Methodology}
\label{ss_methodology}

\begin{table*}
\begin{center}
\begin{tabular}{ |c|c|c|c|c| }
\hline
Number of nodes & Setting & \texttt{build} and \texttt{push}  & \texttt{fetchTransaction} & Total decrease \\
\hline
\multirow{2}{4em}{3 nodes} & Integrated & 65$\pm3$ ms &  \xmark & \multirow{2}{4em}{94.12 \%} \\

& Remote & 1020$\pm21$ ms & 85 ms & \\
\hline
\multirow{2}{4em}{5 nodes} & Integrated & 61$\pm3$ ms & \xmark & \multirow{2}{4em}{93.59 \%} \\

& Remote & 868$\pm38$ ms & 84 ms &\\
\hline
\multirow{2}{4em}{7 nodes}

& Integrated & 66$\pm3$ ms & \xmark & \multirow{2}{4em}{92.31 \%}  \\

& Remote & 773$\pm16$ ms & 86 ms & \\
\hline
\multirow{2}{4em}{10 nodes}
& Integrated & 58$\pm3$ ms & \xmark & \multirow{2}{4em}{93.77 \%}\\
& Remote & 845 ms$\pm32 ms$ & 86 ms & \\
\hline
\end{tabular}
\captionof{table}{Measures of transaction time (averages) for each setting with different networks sizes}
\label{tab:measures}
\end{center}
\end{table*}

In the tests, only the \texttt{build}, \texttt{push} and \texttt{fetchTransaction} calls will be detailed (cf. Section \ref{S_illustrative_scenario}). These three calls will be referred to as \emph{transaction time} in the following. Our first guess would be to include
the \texttt{checkPolicy} call as well. However, figures show that in a scenario where 1000 rules
policy is configured by the users (which is unrealistic), an average 7 ms over 1500 samples
is reached, which is negligible with regard to the measured transaction time. Other calls are lightweight when measured.
Let's now detail why the three calls \texttt{build}, \texttt{push} and \texttt{fetchTransaction} are time-consuming. To build and push a transaction, the node needs to compute a proof of work and check that the transaction is consistent with its state of the ledger. The \emph{fetchTransaction} call is time-consuming due to several varying parameters: 1) the node may be busy processing transactions; 2) the physical distance between the node and the UCS, without integration; 3) the time needed to query the node's ledger, which increases with the number of simultaneous requests.

\textbf{Transaction time.}
To demonstrate actual performance improvements, we measure the time needed for a transaction to be validated and pushed to the network, and the time to fetch the transaction from an IOTA node. These operations are respectively the calls \texttt{build}, \texttt{push} and \texttt{fetchTransaction} of the sequence diagram of Figure \ref{F_sequence}. The \texttt{build} and \texttt{push} calls are only one API call in the Java IOTA library, hence the common measure.
 Tests are conducted in two different configurations, respectively the remote and the integrated settings: (1) the UCS interacting with a remote IOTA node run by an AWS instance; (2) an integrated node running both the UCS and an IOTA node simultaneously, as illustrated in Figure \ref{F_peripheral_vs_integrated}.
   For each test, 1500 samples ($N=1500$) are used and confidence intervals are given with a 95\% confidence level.
 
\subsection{Results}
\label{ss_results}






The resulting experimental measurements are summarized in Table~\ref{tab:measures}. For convenience, only the experiments with 5 nodes $N=5$ will first be discussed. Then, we will detail the impact of the number of nodes on the results and provide explanations.

\textbf{Remote setting}. When the UCS interacts with a remote node, the \texttt{build} and \texttt{push} calls need an average $\overline{t}_{transaction, r} =868\pm 31 ms$ with 5 nodes in the network. The minimum time needed to build and push a transaction is $m_{r}=646 ms$ and the maximum is $M_{r}=7649 ms$. This wide range is due to nodes being often desynchronized - a normal behavior - and having to update the state of the ledger before pushing their own transactions.
Additionally, the UCS has to recover the result of the transaction requiring an additional $\overline{t}_{fetch, r} = 84\pm0.38ms$.

\textbf{Integrated setting}. When using an integrated node on the UCS device, the transaction time gains are substantial. First, the \texttt{fetchTransaction} call does not involve any network calls, and requires less than 1 ms to be completed. Second, our own transactions are prioritized on our own node, substantially decreasing the time to complete the \texttt{build} call. The average time needed to \texttt{build} and \texttt{push} a transaction plummets to $\overline{t}_{transaction, l}=61\pm2.86ms$.

\textbf{Number of nodes.} As the number of nodes increases, the impact of integration in percentage remains stable, ranging from 92.31\% to 94.12\% transaction time decrease. It means that on average, the transaction time in our integration scheme is 13 to 18 times faster compared to the remote node model.

The time to build and push transactions decreases before 7 nodes, then start to go up when using 10 nodes. Firstly, the number of messages spammed increases with the number of nodes. We can therefore observe the expected behavior of IOTA: as the number of transactions per second increases, transactions are validated faster. Secondly, when reaching 10 nodes, there are approximately 50 messages spammed per second as each node spams 5 messages per second. There could be two explanations for this behavior. First, the network could saturate due to the Coordinator milestones validation. This issue is well-identified by both the IOTA foundation and academics \cite{Wang2022, Conti2022} and is the motivation for the development of IOTA 2.0, willing to remove the Coordinator. However, the Coordinator runs on the same device in both integrated and remote settings, but the transaction time does not increase for 10 nodes in the integrated setting. Therefore, we can conclude that it is not the Coordinator that saturates, but the remote node used for the testing. Indeed, the nodes running on AWS are resource-constrained, and processing 50 transactions per second is time-consuming for an instance with 1 Gio of RAM and 1 vCPU.


\section{Privacy evaluation}
\label{S_privacy_evaluation}
 In this section, we conduct a privacy threat analysis, to highlight the risks for the users in the electricity consumption prediction scenario (Section \ref{S_illustrative_scenario}). Based on this analysis, we then detail how usage control mitigates several privacy risks.

\subsection{Threat model} The \emph{usage control system} is considered \emph{honest-but-curious}. It fulfills its usage control tasks, but may be interested in collecting undue data about the users it monitors. This behavior may be financially motivated, i.e., to sell the users' data afterward. In particular, the UCS has interest in gathering the system calls and network calls of the monitored users, both carrying valuable, privacy-sensitive data.

\emph{Data buyers} and \emph{data providers} are considered honest-but-curious as well. Notably, data buyers try to infer more data from the protected data they buy, but do not try to bypass the UCS monitoring or to compromise the database.
The \emph{Cassandra distributed database} is considered trusted and not compromised.

\subsection{Privacy risks - Illustrative scenario}
\label{ss_privacy_risks}

The privacy threat assessment is conducted on the energy consumption prediction scenario (Section \ref{S_illustrative_scenario}), to describe the threats more accurately. The analysis can nonetheless be generalized to any use case requiring to buy and sell data. The \emph{LINDDUN privacy threat modeling framework} \cite{Deng2011} is used to describe extensively the potential threats to privacy and help to frame the analysis.

\textbf{Items of interest.} Items of interest (IOI) in LINDDUN refer to any data element, including users' personal data or transaction data, that is considered privacy-sensitive. It is paramount to exhaustively identify which items are of interest before conducting the privacy threat assessment, to ensure the full listing of the threats. Items of interest can be subjects, messages, actions or data. In the electricity consumption prediction scenario:
\begin{itemize}
    \item the temperature and the humidity data, which are protected data. Protected data are geolocated;
    \item the transaction data, including but not limited to users' addresses and transaction values;
    \item users' data, i.e., data buyer and data provider personal data;
    \item usage control data, such as the results of access requests;
    \item metadata, such as the time the data was created or added to the Cassandra database, that can be used to infer other data.
    \item data buyers, data providers, the UCS and the Cassandra database, i.e, subjects;
    \item network messages between the subjects, and the messages between the UCS components (cf. Section \ref{ss_usage_control});
    \item actions carried out by the subjects, such as an access request by the data buyer, entry insertion on the database by the data provider or the beginning of an access request evaluation by the UCS.
\end{itemize}

In the following, transaction data are ignored, as preserving privacy in distributed ledger is a specific, orthogonal research topic. Legal compliance in particular is troublesome for distributed ledgers \cite{Haque2021}, and privacy-enhancing technologies have been designed to address privacy threats, such as cryptocurrency mixers for linkability \cite{Sarfraz2019}, \cite{Glaeser2022}.

\textbf{Linking (L).} Linking occurs when two \emph{items of interest} are associated to learn more about an individual or a group. Due to the diversity of agents and technologies involved, linking threats are numerous in the electricity consumption prediction scenario:

\begin{enumerate}
    \item linkage between a protected data and its owner (user data). The location in the temperature or humidity data can be used to facilitate re-identification of the data providers (I);
    \item linkage between a data buyer and a data provider. It simplifies re-identification when one of them is identified (I);
    \item linkage between metadata and usage control data, as both are written on the ledger.
\end{enumerate}

Linking leads to other identifying (I), detection (D) and non-repudiation (N) threats and is triggered by data disclosure (D) - the more data available, the more likely are linking threats.

\textbf{Identifying (I).} Identification is occurring when an attacker learns the identity of an individual, breaking its anonymity or pseudonymity. This threat distinguishes \emph{identified data} where the identity is explicitly maintained, and \emph{identifiable data} which enables to derive the identity indirectly. Users' attributes, e.g., IP address or processed by the usage control system to make an access decision can be used to infer the user's true identity. In the scenario, the risks are to re-identify data buyers and data providers, using protected data, usage control data or metadata.

\textbf{Non-repudiation (N).} Non-repudiation, in LINDDUN privacy threat assessment, is the ability to attribute a claim to an individual. For example, the impossibility for a data buyer to deny they bought and accessed a protected data, or for data providers to deny they generated a given data, are non-repudiation threats.

\textbf{Detecting (D).} Detecting is the ability to deduce the involvement of an individual in an action with observation. Detecting does not require being able to read the actual data. Knowing that the data exists is enough to infer more sensitive information. Detecting can be done by observing \emph{communications} or \emph{application side-effects}. Detecting threats in the electricity consumption prediction scenario includes: 
\begin{enumerate}
    \item detecting that a user is monitored by analyzing the communications of the UCS. An attacker learns that the user has likely bought valuable data;
    \item detecting what are the protected data and where they are disseminated, without the actual content of the data, by analyzing the UCS logs;
    \item detecting users' attributes by intercepting the communications between the policy information points (PIP) and the context handler (CH) (cf. Section \ref{ss_usage_control});
\end{enumerate}

\textbf{Data disclosure (D).} Data disclosure is the
immoderate collecting, storing, processing or sharing of personal data. This generic threat focuses on four characteristics:
\begin{enumerate}
    \item \emph{unnecessary data types}, if the data granularity or sensitivity is too detailed;
    \item \emph{excessive data volume}, if the amount or the frequency of data processing is too high;
    \item \emph{unnecessary processing}, if the data is processed or disseminated out of necessity;
    \item \emph{excessive exposure}, refers to how widely accessible the data is and to whom the data is shared.
\end{enumerate}

In the energy consumption prediction scenario, data disclosure can occur if a data provider has access to protected data while not fulfilling the policy's conditions, or can disseminate it outside of the UCS monitoring perimeter. Similarly, data disclosure occurs if the UCS collects too detailed data about the data buyers to monitor them.

\textbf{Unawareness (U).} Unawareness occurs when an individual is insufficiently informed and involved in the processing of personal data.
Unawareness occurs in the scenario if the data providers are not informed about the privacy threats they face by
selling their data with associated geolocation. The user may also be unaware of the privacy risks of accepting to be monitored by the UCS, which writes usage control data on the ledger, that are potentially privacy-sensitive (cf. Section \ref{ss_methodology}) 

\textbf{Non compliance (N).} Non-compliance is the deviation from security and data management best practices, standards and legislation. This risk occurs if the processing of any item of interest is considered unlawful based on the specified policies and the applicable regulations, e.g., GDPR \cite{EUdataregulations2018}. In the scenario, the risks are that although the data provider specifies a policy, it is not properly enforced by the usage control system.


\begin{table*}
\centering
\begin{tabular}{ |c|c|c|c| }
\hline
Threat & Scenario example & Mitigated by the UCS\\
\hline
Linking & Link between data owner and buyer & ?\\
Identifying & Data provider is re-identified & ?\\
Non-repudiation & Data provider can not deny data generation & \xmark \\
Detecting & Detect a user is monitored & \cmark\\
Data disclosure & Protected data dissemination & \cmark\\
Unawareness & Inference risks of location unknown  & \cmark\\
Non-compliance & Data provider policy not enforced & \cmark\\
\hline
\end{tabular}
\vspace*{1mm} 
\caption{LINDDUN privacy threat analysis, based on the illustrative scenario}
\label{tab:threats}
\end{table*}

\subsection{Threat mitigation with usage control}
\label{ss_mitigation}

As a privacy-enhancing technology, usage control is designed to address a significant part of the above-mentioned privacy threats. We next detail how these threats are addressed, and which ones can not be mitigated by the UCS. The results are summarized in Table \ref{tab:threats}.

\textbf{Linkability. (?)}
The usage control system has only a partial impact in preventing this threat. Notably, the linkage between metadata and usage control data, that are both unrestricted data available publicly on the ledger, can be accessed without monitoring in the first place. However, the UCS monitors the dissemination and the processing of the protected data, limiting the risks to link them to their owners.

\textbf{Identifying. (?)} Similarly, the usage control system must be able to identify the data buyer to fulfill its mission. While it can prevent users from disclosing \emph{identified data}, it is much harder for it to prevent inference from \emph{identifiable data}.

\textbf{Non-repudiation. \xmark}  To fulfill its task, the UCS needs to ensure that a monitored data buyer can not decline having disseminated the data, or having processed it in an unlawful manner. Therefore, the UCS not only does not guarantee non-repudiation, but also writes usage control data on the distributed ledger, making non-repudiation impossible.

\textbf{Detecting. \cmark} To mitigate this threat, all communications are secured using TLS, notably the communication between the usage control system and the users, the context handler and the PIPs, as well as the context handler and the PEPs. Besides, unless the UCS is compromised, an attacker does not have access to the logs of the UCS.

\textbf{Data disclosure. \cmark} The data processing is monitored by the usage control system, mitigating the \emph{excessive data volume} threat. The usage control system, by monitoring both the information flow and the usage of the data, also prevents \emph{unnecessary processing}. \emph{Excessive exposure} is prevented as part of the access control to the data.

\textbf{Unawareness. \cmark} The usage control system asks data owners to design data policies themselves, directly addressing the unawareness threat.

\textbf{Non-compliance. \cmark} The usage control system monitors the data buyers, stopping them from processing the data unlawfully. The usage control system, considered honest-but-curious, enforces the proper usage control policy specified by the data provider.

Usage control addresses four categories of threats directly, i.e., detecting, data disclosure, unawareness and non-compliance. Due to the diversity of the data and agents considered, linking and identifying threats are only partially mitigated. Only the non-repudiation threat is not considered by usage control.
\section{Conclusion}
\label{S_conclusion}

The Internet of Things has simultaneous requirements in terms of performance and privacy. Distributed ledger technologies are interesting for the IoT because they reduce the cost and decentralize the network. While blockchains are mostly not fitted for the Internet of Things, DAG-based distributed ledgers conversely provide several properties of interest: free or very cheap transactions, high throughput and disintermediation. Usage control enables users to monitor their data and provides the technical basis to enforce their policies. After identifying directed acyclic graphs as a relevant technology, we proposed a solution using both IOTA, a DAG-based distributed ledger, and usage control. The originality of the solution is to integrate the usage control system into the IOTA network, instead of using it as an external tool.

Starting from an illustrative scenario about the prediction of electricity consumption using temperature and humidity data, a proof-of-concept has been implemented to demonstrate the efficiency of the integration.
The performance tests have shown that the time needed for UCS transactions decreases by \emph{more than 90\%} when the UCS is integrated into the private network as a node.
A privacy threat assessment was conducted with the LINDDUN framework, highlighting the privacy benefits of usage control. Usage control addresses most of the LINDUN privacy threats, completely (4 out of 7) or partially (6 out of 7), excluding non-repudiation.

While this article focuses on usage control, the integration principle can be generalized to any system that requires to process distributed ledger transactions and that can be deployed on a node. The network security will always benefit from having more nodes (cf. Section \ref{S_integration}), and throughput will also increase for directed acyclic graphs. Similarly, a device will benefit from deploying a node if it has to analyze the transaction, which is the case for the UCS so as to monitor access.

\section*{Data availability statement}
All the code as well as evaluation results are available in a repository (\url{https://zenodo.org/badge/latestdoi/567207401}), notably to reproduce the tests. The settings of the AWS instances, including the software versions, are detailed on the repository.

\section*{Acknowledgments}
This paper is supported by the Future \& Ruptures program of Fondation Mines-Télécom, the Institut Mines-Télécom VP-IP Chair on Values and Policies of Personal Information (\url{https://cvpip.wp.imt.fr}) and Energy4Climate (E4C) interdisciplinary center (\url{https://www.e4c.ip-paris.fr/})

\bibliographystyle{plain}
\bibliography{bibliography}

\end{document}